\documentclass[12pt]{article}

\newcommand{\di}{\partial}
\newcommand{\be}{\begin{displaymath}}
\newcommand{\ee}{\end{displaymath}}
\newcommand{\benumber}{\begin{equation}}
\newcommand{\eenumber}{\end{equation}}
\newcommand{\eref}[1]{(\ref{#1})}
\newcommand{\grad}{\nabla}

\newcommand{\half}{\frac{1}{2}}

\newcommand{\bv}[1]{\mbox{\boldmath${#1}$}}

\newcounter{saveeqn}%
\newcommand{\alpheqn}{\setcounter{saveeqn}{\value{equation}}%
\stepcounter{saveeqn}\setcounter{equation}{0}%
\renewcommand{\theequation}
	{\mbox{\arabic{saveeqn}\alph{equation}}}}%
\newcommand{\reseteqn}{\setcounter{equation}{\value{saveeqn}}%
\renewcommand{\theequation}{\arabic{equation}}}%

\newcommand{\mainlabel}[1]{\renewcommand{\theequation}{\arabic{saveeqn}}%
\label{#1}
\renewcommand{\theequation}{\mbox{\arabic{saveeqn}\alpha{equation}}}}%

\title{On the Asymptotic Behaviour of Cosmological Models in Scalar-Tensor 
Theories of Gravity}

\author{ \bf Andrew Billyard \thanks{\small {Department of
Mathematics, Statistics and Computing Science and Department of
Physics, Dalhousie University, Halifax, Nova Scotia, Canada B3H 3J5}}
\and \bf Alan Coley$^{ *}$ \and \bf Jesus Ib\'{a}\~{n}ez \thanks{\small
{Dpto. Fisica Teorica, Universidad del Pais Vasco, Bilbao, Spain.}}}

\begin{document}

\bibliographystyle{aip}

\maketitle

\begin{abstract}
We study the qualitative properties of cosmological models in
scalar-tensor theories of gravity by exploiting the formal equivalence
of these theories with general relativity minimally coupled to a
scalar field under a conformal transformation and field redefinition.
In particular, we investigate the asymptotic behaviour of
spatially homogeneous cosmological models in a class of scalar-tensor
theories which are conformally equivalent to general relativistic Bianchi
cosmologies with a scalar field and an exponential potential whose
qualitative features have been studied previously.  Particular attention
is focussed on those scalar-tensor theory cosmological models, which
are shown to be self-similar, that correspond to general relativistic
models that play an important r\^{o}le in describing the asymptotic
behaviour of more general models (e.g., those cosmological models that
act as early-time and late-time attractors).
\end{abstract}

\section{\bf Introduction}

\

Scalar-tensor theories of gravity are currently of great interest,
partially due to the fact that such theories occur as the low-energy
limit in superstring theory (see \cite{Horowitz1990a} and references
therein).  The first scalar-tensor theories to appear (with
$\omega=\omega_0$) were due to Jordan \cite{Jordan1949a,Jordan1959a},
Fierz \cite{Fierz1956a} and Brans and Dicke \cite{Brans1961a} and 
the most general
scalar-tensor theories were formulated by Bergmann \cite{Bergmann1968a}, 
Nordtvedt \cite{Nordtvedt1970a} 
and Wagoner \cite{Wagoner1970a}.  The observational limits on scalar-tensor
theories include solar system tests \cite{Buchmann1996a, 
Abramovici1992a,Hough1992a,Bradaschia1990a} and 
cosmological tests such as Big Bang nucleosynthesis constraints 
\cite{Barrow1987a,Serna1996b}.

The possible isotropization of spatially homogeneous cosmological
models in scalar-tensor theories has been studied previously.  For
example, Chauvet and Cervantes-Cota \cite{Chauvet1995a} have studied
the possible isotropization of Bianchi models of types $I$, $V$ and
$I\!X$ within the context of Brans-Dicke theory without a
scalar potential, but with barotropic matter, $p=(\gamma-1)\mu$, by
studying exact solutions at late times.  Mimoso and Wands
\cite{Mimoso1995a} have studied Brans-Dicke theory with a variable
$\omega=\omega(\phi)$ in the presence of barotropic matter (but with
no scalar field potential) and, in particular, gave forms for $\omega$
under which Bianchi $I$ models isotropize.  We note that there is a formal
equivalence between such a theory (with $\gamma\neq 2$) and
a scalar-tensor theory with a potential but without matter, via the
field redefinitions $V\equiv(2-\gamma)\mu$ and $\omega
\grad_a\phi\grad_b\phi \rightarrow \omega
\grad_a\phi\grad_b\phi -\gamma\mu \phi \delta^0_a\delta^0_b$. 

In a recent paper \cite{Coley1997a} (see also \cite{Ibanez1995a} and
\cite{vandenHoogen1997a}), cosmological models containing a scalar
field with an exponential potential were studied.  In particular, the
asymptotic properties of the spatially homogeneous Bianchi models, and
especially their possible isotropization and inflation, were
investigated.  Part of the motivation for studying such models is that
they can arise naturally in alternative theories of gravity
\cite{Burd1988a}; for example, Halliwell \cite{Halliwell1987a} has shown that
the dimensional reduction of higher-dimensional cosmologies leads to
an effective four-dimensional theory coupled to a scalar field with an
exponential self-interacting potential.  

The action for
the general class of scalar-tensor theories (in the so-called
Jordan frame) is given by \cite{Bergmann1968a,Wagoner1970a},
\benumber S=
\int \sqrt{-g} \left[ \phi R - \frac{\omega(\phi)}{\phi}g^{ab}\phi_{,a}\phi_{,b}-2V(\phi) + 2{\cal{L}}_m \right] d^4x. \label{bdaction}
\eenumber
However, under the conformal transformation and field redefinition
\cite{Barrow1994a, Liddle1992a, Mimoso1995a}
\alpheqn
\begin{eqnarray}
\mainlabel{trans}
 g^*_{ab} & = & \phi g_{ab} \label{trans1} \\
\frac{d\varphi}{d\phi} & = & 
      \frac{\pm\sqrt{\omega(\phi)+3/2}}{\phi}, \label{trans2}
\end{eqnarray}
\reseteqn
the action becomes (in the so-called Einstein frame)
\benumber
S^*=\int \sqrt{-g^*}\left[ R^*-g^{*ab}
\varphi_{,a}\varphi_{,b}-2\frac{V(\phi)}{\phi^2} + 2\frac{
{\cal{L}}_m}{ \phi^2} \right] d^4x, \label{eaction}
\eenumber
which is the action for general relativity (GR) containing a scalar field
$\varphi$ with the potential
\benumber
V^*(\varphi)= \frac{V(\phi(\varphi))}{\phi^2(\varphi)}. \label{vstar}
\eenumber

Our aim here is to exploit the results in previous work
\cite{Coley1997a} to study the asymptotic properties of scalar-tensor
theories of gravity with action \eref{bdaction} which under the
transformations \eref{trans} transform to general relativity with a
scalar field with the exponential potential given by
\benumber
V^* = V_0 e^{k\varphi} \label{exp},
\eenumber
where $V_0$ and $k$ are positive constants.  That is, since we
know the asymptotic behaviour of spatially homogeneous Bianchi models
with action \eref{eaction} with the exponential potential \eref{exp},
we can deduce the asymptotic properties of the corresponding
scalar-tensor theories under the transformations
\eref{trans}\footnote{The possible isotropization of spatially
homogeneous scalar-tensor theories which get transformed to a model
with an effective potential which passes through the origin and is
concave up may be deduced from the results of Heusler
\cite{Heusler1991a}} (so long as the transformations are not
singular!).  In particular, we are concerned with the possible
isotropization and inflation of such scalar-tensor theories.

The outline of the paper is as follows.  In section 2, we review the 
framework within which GR and a scalar field with a potential (Einstein 
frame) is formally equivalent to a scalar-tensor theory with a potential 
(Jordan frame), concentrating on both the exact and approximate forms for 
the parameters $V$ and $\omega$ in the Jordan frame.  In particular, we 
discuss the explicit example of the Brans-Dicke theory with a power-law 
potential and we also discuss the conditions which lead to 
appropriate late-time behaviour as dictated by solar system and 
cosmological tests.  In section 3, we then apply the conformal 
transformations to Bianchi models studied in the Einstein frame to 
produce exact solutions which represent the asymptotic behaviour of more 
general spatially-homogeneous models in the Jordan frame (for 
$\omega=\omega_0$, a constant).  
These Brans-Dicke models are self-similar and the corresponding homothetic 
vectors are also exhibited.  We conclude with a discussion in section 4.

\section{\bf Analysis}

For scalar field Bianchi models the conformal factor in \eref{trans1}
is a function of $t$ only (i.e., $\phi=\phi(t)$), and hence under
(non-singular) transformations \eref{trans} the
Bianchi type of the underlying models does not change (i.e., the
metrics $g_{ab}$ and $g^*_{ab}$ admit three space-like Killing vectors
acting transitively with the same group structure).  In general, in
the class of scalar-tensor theories represented by \eref{bdaction}
there are two arbitrary (coupling) functions $\omega(\phi)$ and
$V(\phi)$.  The models which transform under \eref{trans} 
to an exponential potential model, in which the two
arbitrary functions $\omega$ and $V$ are constrained by \eref{trans2}
and \eref{vstar}, viz., 
\benumber \frac{\phi}{V}\frac{dV}{d\phi}=2 \pm k
\sqrt{\frac{3}{2}+\omega(\phi)} ,\label{Vdefine}
\eenumber
is a special subclass with essentially one arbitrary function.
Although only a subclass of models obey this constraint, this subclass
is no less general than massless scalar field models ($V=0$; see, for example
\cite{Mimoso1995a}) or Brans-Dicke models with a potential
($\omega=\omega_0$, constant), which are often studied in the
literature.  Indeed, the asymptotic analysis in this paper is valid
not only for ``exact'' exponential models, but also for scalar-tensor
models which transform under \eref{trans} to a
model in which the effective potential is a linear combination of
terms involving exponentials in which the dominant term asymptotically
is a leading exponential term; hence the analysis here is rather more
general (we shall return to this in the next section).  For the
remainder of this paper we shall not explicitly consider ordinary
matter; i.e., we shall set the matter Lagrangians in \eref{bdaction} and
\eref{eaction} to zero.  Matter can be included in a straightforward
way \cite{Heusler1991a, Mimoso1995a, Kitada1992a}.

\subsection{\bf Exact Exponential Potential Models \label{exact}}

Scalar-tensor models which transform under \eref{trans} to a model with an 
exact exponential potential satisfy
equations \eref{trans2} and \eref{vstar} with \eref{exp}, viz.,
\begin{eqnarray}
\frac{d\varphi}{d\phi} & = & \pm \frac{\sqrt{\omega(\phi)+3/2}}
{\phi} \\
V_0 e^{k\varphi} & = & \frac{V(\phi)}{\phi^2}.
\end{eqnarray}
So long as the transformations \eref{trans} remain non-singular we can
determine the asymptotic properties of the underlying scalar-tensor
theories from the asymptotic properties of the exact exponential
potential model.  These properties were studied in \cite{Coley1997a}.
We recall that the asymptotic behaviour depends crucially on the
parameter $k$ (in \eref{exp}) which will be related to the various
physical parameters in the scalar-tensor theory \eref{bdaction}.

In particular, in \cite{Coley1997a} it was shown that all scalar field
Bianchi models with an exponential potential \eref{exp} (except a
subclass of the Bianchi type $I\!X$ models which recollapse) isotropize to the future
if $k^2\leq 2$ and, furthermore, inflate if $k^2<2$; if $k=0$ these
models inflate towards the de Sitter solution and in all other cases
they experience power-law inflationary behaviour.  If $k^2 > 2$, then
the models cannot inflate, and can only isotropize to the future if
the Bianchi model is of type $I$, $V$, $V\!I\!I$, or $I\!X$.  Those
models that do not isotropize typically asymptote towards a
Feinstein-Ib\'{a}\~{n}ez anisotropic model \cite{Feinstein1993a}.
Bianchi $V\!I\!I_h$ models with $k^2>2$ can indeed isotropize
\cite{Coley1997a} but do not inflate, while generically the 
ever-expanding Bianchi $I\!X$ models do not isotropize \cite{vandenHoogen1998a}.

Therefore, at late times and for each specific choice of
$\omega(\phi)$ both the asymptotic behaviour of the models and
the character of the conformal transformation \eref{trans} may be
determined by the behaviour of the scalar field $\varphi$ at the
equilibrium points of the system in the Einstein frame. Recently this
behaviour has been thoroughly investigated \cite{Coley1997a}. We shall
summarize only those aspects relevant to our study.  The existence of
GR as an asymptotic limit at late times is also determined by the asymptotic
behaviour of the scalar field; we shall return to this issue in section
\ref{gr}.

For spatially homogeneous space-times the scalar field $\varphi$ is
formally equivalent to a perfect fluid, and so expansion-normalized
variables can be used to study the asymptotic behaviour of Bianchi
models \cite{Coley1997a,Wainwright1997a}.  The scalar field variable,
$\Psi$, is defined by
\begin{equation}
\Psi\equiv \frac{\dot\varphi}{\sqrt{6} \theta^{*}} \label{jesus1},
\end{equation}
where $\theta^{*}$ is the expansion of the timelike congruences
orthogonal to the surfaces of homoge\-neity\footnote{Note that
$\theta^*>0$ for all Bianchi models except those of type $I\!X$.}. At
the finite equilibrium points of the reduced system of autonomous
ordinary differential equations, where $\Psi$ is a finite constant, it
has been shown \cite{Wainwright1997a} that
$\theta^{*}=\theta^{*}_0/t^{*}$, where $t^{*}$ is the time defined in
the Einstein frame:
\begin{equation}
dt^{*}=\pm \sqrt{\phi}\, dt. \label{jesus2}
\end{equation}
From equation  \eref{jesus1} it follows that $\dot\varphi \propto 1/t^{*}$,
whence upon substitution into the Klein-Gordon equation
\begin{equation}
\ddot\varphi+\theta^{*}\dot\varphi+{\partial V^{*}\over \partial \varphi}=0,
\label{jesus3}
\end{equation}
we find that at the finite equilibrium points
\begin{equation}
\varphi (t^{*})=\varphi_0-{2\over k}\ln t^{*};\qquad k\neq 0,
\label{jesus4}
\end{equation}
where $\varphi_0$ is a constant. Hence, from equation \eref{trans2} we
can obtain $\phi$ as a function of $t^{*}$, provided a particular
$\omega(\phi)$ is given.  From equation \eref{jesus2} we can then find the
relationship between $t^{*}$ and $t$, and consequently obtain $\phi$
as a function of $t$, and hence determine the asymptotic behaviour of
$\phi(t)$ for a given theory with specific $\omega(\phi)$ (in the
Jordan frame).  Specifically, we can determine the
possible isotropization and inflation of a given scalar-tensor theory
in a very straightforward way.

As mentioned above, the behaviour determined from the key equation
\eref{jesus4} is not necessarily valid for all Bianchi models.  For
the Bianchi models in which the phase space is compact, the 
equilibrium points represent models that do have the behaviour described by
\eref{jesus4}, as do the finite equilibrium points in Bianchi models with
non-compact phase spaces.  It is possible that the infinite equilibrium
points in these non-compact phase spaces also share this behaviour,
although this has not been proven.  Finally, from equations \eref{trans}
we note that since the asymptotic behaviour is governed by
\eref{jesus4}, the corresponding transformations are non-singular and
this technique for studying the asymptotic properties of spatially
homogeneous scalar-tensor theories is valid.

\subsection{\bf An Example}

Suppose we consider a Brans-Dicke theory with a power-law potential, viz.,
\begin{eqnarray}
\omega(\phi) & = & \omega_0 \label{bdw} \\
V & = & \beta \phi^\alpha \label{power}
\end{eqnarray}
(where $\beta$ and $\alpha$ are positive constants), then \eref{trans2} integrates to yield \benumber
\phi=\phi_0 \ \exp\left(\frac{\varphi-\varphi_0}{\bar{\omega}}\right), \label{powerphi} 
\eenumber where \benumber
\bar{\omega}\equiv\pm\sqrt{\omega_0+3/2},\label{barw} \eenumber and hence
\eref{vstar} yields
\benumber
V^* = V_0 e^{\bar{k}\varphi},
\eenumber
where the critical parameter $\bar{k}$ is given by 
\benumber \bar{k}=\frac{\alpha-2}{\bar{\omega}}. \eenumber
From \cite{Coley1997a} we can now determine the asymptotic behaviour of
the models in the Einstein frame, as discussed in
section \ref{exact}, for a given model with specific values of $\alpha$
and $\bar{\omega}$ (and hence a particular value for $\bar{k}$).

The possible isotropization of the given scalar-tensor theory can now
be obtained directly (essentially by reading off from the proceeding
results - see subsection 3.1).  For example, the inflationary
behaviour of the theory can be determined from equations
\eref{trans1}, \eref{jesus2} and \eref{jesus4}.  Let us further
discuss the asymptotic behaviour of the corresponding scalar-tensor
theories (in the Jordan frame).  From equations
\eref{jesus2}, \eref{jesus4} and \eref{powerphi} we have that
asymptotically
\benumber
\phi =\tilde{\phi}_0 \left[\pm (t-t_0) \left(1+\frac{1}{k\bar{\omega}}\right)\right]^{-2/(1+k\bar{\omega})},
\eenumber
where the $\pm$ sign is determined from \eref{jesus2}.  Both this sign
and the signs of $\bar{\omega}$ and $1+k\bar{\omega}$ are crucial in
determining the relationship between $t^*$ and $t$; i.e., as $t^*
\rightarrow \infty$ either $t\rightarrow \pm \infty$ or $t\rightarrow
t_0$ and hence either $\phi \rightarrow 0$ or $\phi \rightarrow
\infty$, respectively, as $\varphi \rightarrow -\infty$.

\subsubsection{\bf A Generalization}

Suppose again that $\omega=\omega_0$, so that \eref{powerphi} also
follows, but now $V$ is a sum of power-law terms of the form
\benumber
V = \sum^m_{n=0} \beta_n \phi^{\alpha_n},
\eenumber
where $m>1$ is a positive integer.  Then \eref{vstar} becomes
\begin{eqnarray}
\nonumber
V^* & = & \sum^m_{n=0} \beta_n \phi^{\alpha_n-2} \\
    & = & \sum^m_{n=0} \bar{\beta}_n \exp(\bar{k}_n \varphi); \ \ \ 
\bar{k}_n = 
            \frac{\alpha_n-2}{\bar{\omega}}.
\end{eqnarray}
For example, if 
\be
V = V_0 + \half m \phi^2 + \lambda \phi^4,
\ee
then
\be
V^* = \bar{V}_0 e^{-2\varphi/\bar{\omega}} + \half \bar{m} 
    + \bar{\lambda}e^{2\varphi/\bar{\omega}}
\ee
(with obvious definitions for the new constants), which is a linear
sum of exponential potentials.  Asymptotically one of these potentials
will dominate (e.g., as $\varphi \rightarrow +\infty$, $V^* \rightarrow
\bar{\lambda}e^{2\varphi/\bar{\omega}}$) and the asymptotic properties can
be deduced as in the previous section.

\subsubsection{\bf Approximate Forms}

In the last subsection we commented upon the asymptotic properties of
a scalar-tensor theory with the forms for $\omega$ and $V$ given by
\eref{bdw} and \eref{power}.  Let us now consider a scalar-tensor
theory with forms for $\omega$ and $V$ which are approximately given
by \eref{bdw} and \eref{power} (asymptotically in some well-defined sense)
in order to discuss whether both theories will have the same
asymptotic properties.  In doing so, we hope to determine whether the
techniques discussed in this paper have a broader applicability.

We assume that $\omega$ and $V$ are analytic at the asymptotic values
of the scalar field in the Jordan frame in an attempt to
determine whether their values correspond to the appropriate forms for
$\varphi$ and $V^*$ in the Einstein frame, namely whether
$\varphi\rightarrow -\infty$ and the leading term in $V^*$ is of the
form $e^{k\varphi}$.

Consider an analytic expansion for $\phi$ about $\phi=0$:
\begin{eqnarray}
\omega & = & \sum_{n=0}^{\infty}\omega_n \phi^n \\ V & = &
\sum_{n=0}^{\infty} V_n \phi^n,
\end{eqnarray}
where all $\omega_n$ and $V_n$ are constants.  Using \eref{trans} we
find, up to leading order in $\phi$, that for $\omega_0 \neq -3/2$
\benumber \varphi-\varphi_0 \approx \bar{\omega} \ln\phi, \eenumber so
that $\varphi \rightarrow \pm \infty$ (depending on the sign in
\eref{barw}) for $\phi\rightarrow 0$.  The potential in the Einstein
frame is (to leading order) \benumber V^* \approx \exp\left\{
-\frac{2(\varphi-\varphi_0)}{\bar{\omega}} \right\}.  \eenumber Hence,
the parameter $k$ of \eref{exp} is defined here as $k\equiv
-2/\bar{\omega}$.  For $\omega_0=-3/2$ we have
\begin{eqnarray}
(\varphi-\varphi_0)^2 & \approx & 4 \omega_1 \phi \\ V^* & \approx &
\frac{16 \omega_1^2}{(\varphi-\varphi_0)^4},
\end{eqnarray}
so that $\varphi \not{\!\!\rightarrow} -\infty$ as $\phi\rightarrow
0$.

Next, let us consider an expansion in $1/\phi$, valid for
$\phi\rightarrow \infty$:
\begin{eqnarray}
\omega & = & \sum_{n=0}^{\infty}\frac{\omega_n}{ \phi^n}, \\
V & = & \sum_{n=0}^{\infty} \frac{V_n }{\phi^n}.
\end{eqnarray}
For $\omega_0\neq -3/2$, the results are similar to the $\phi=0$ expansion:
\begin{eqnarray}
\varphi-\varphi_0 & \approx & - \bar{\omega} \ln\phi \\
V^* & \approx & \exp\left\{ \frac{2(\varphi-\varphi_0)}{\bar{\omega}} \right\},
\end{eqnarray}
where now $\varphi \rightarrow \mp \infty$ as $\phi\rightarrow \infty$.
When $\omega_0=-3/2$, we obtain
\begin{eqnarray}
(\varphi-\varphi_0)^2 & \approx & \frac{4 \omega_1}{ \phi} \\
V^* & \approx & \frac{(\varphi-\varphi_0)^4}{16 \omega_1^2}.
\end{eqnarray}
It is apparent that the sign of $\bar{\omega}$ is important in
determining whether $\phi\rightarrow \infty$ or $\phi\rightarrow 0$ in order 
to obtain the appropriate form for $\varphi$, as was
exemplified at the end of section 2.2.

Finally, in the event that $\omega$ and $V$ are analytic about some
finite value of $\phi$, namely $\phi_0$, it can be shown that $\varphi
\rightarrow \varphi_0$ as $\phi \rightarrow \phi_0$.  Hence, if one
insists that $\omega$ remain analytic as $\omega\rightarrow\omega_0$ in
the limit of $\varphi\rightarrow -\infty$, then $\phi$ must either
vanish or diverge, and the GR limit is not obtained.  This would
then suggest that if one imposed $\varphi\rightarrow -\infty$ for
$\phi \rightarrow \phi_0$ then $\omega$ would not be analytic about
$\phi=\phi_0$.

\subsection{\bf Constraints on Possible Late-Time Behaviour\label{gr}}
In this paper we are concerned with the possible asymptotic behaviour
of cosmological models in scalar-tensor theories of gravity.  However,
there are physical constraints on acceptable late-time behaviour (as $t^*
\rightarrow \infty$; see equation \eref{jesus2}).  For example, such
theories ought to have GR as an asymptotic limit at late times (e.g.,
$\omega\rightarrow \infty$ and $\phi\rightarrow \phi_0$) in order for
the theories to concur with observations such as solar system tests.
In addition, cosmological models must `isotropize' in order to be in
accord with cosmological observations.

Nordtvedt \cite{Nordtvedt1970a} has shown that for scalar-tensor
theories with no potential, $\omega(\phi)\rightarrow\infty$ and
$\omega^{-3}d\omega/d\phi \rightarrow 0$ as $t\rightarrow \infty$ in
order for GR to be obtained in the weak-field limit. Similar
requirements for general scalar-tensor theories with a non-zero
potential are not known, and as will be demonstrated from the
consideration of two particular examples found in the literature, not
all theories will have a GR limit.  

The first example is the
Brans-Dicke theory ($\omega=\omega_0=constant$) with a power-law
self-interacting potential given by \eref{power} studied earlier in
subsection 2.2. In this case, $\phi$ is given by equations
\eref{powerphi} and \eref{barw} and the potential is given by
\eref{power}, viz.,
\be
V(\phi)=\beta\phi^\alpha; \qquad \alpha=2\mp k \sqrt{\omega_0+3/2}.
\ee
The $\alpha=1$ case for FRW metrics was studied by Kolitch
\cite{Kolitch1996a} and the $\alpha=2$ ($k=0$) case, corresponding to a
cosmological constant in the Einstein-frame, was considered for FRW
metrics by Santos and Gregory \cite{Santos1996a}.  Earlier we
considered whether anisotropic models in Brans-Dicke theory with a
potential given by \eref{power} will isotropize. Assuming a large
value for $\omega_0$, as suggested by solar system experiments, we
conclude that for a wide range of values for $\alpha$ the models
isotropize. However, in the low-energy limit of string theory where 
$\omega_0=-1$ the models are only guaranteed to isotropize for $1<\alpha<3$.

Substituting \eref{jesus4} in \eref{powerphi} we get 
\begin{equation}
\phi\sim (t^{*})^{\pm 2\delta}, \qquad \delta={1\over k}\;\sqrt{{2\over 
3+2\omega_0}}.
\label{jesus8}
\end{equation}
Now, substituting the above expression into equation \eref{jesus1}, we obtain
$t^{*}$ as a function of $t$ and hence we obtain
\begin{equation}
\phi\sim t^{{\pm 2\delta\over 1\mp\delta}}.
\label{jesus9}
\end{equation}
Depending on the sign, we deduce from this expression that for large
$t$ the scalar field tends either to zero or to infinity and so
this theory, with the potential given by \eref{power}, does not have a
GR limit.

In the second example we assume that
\begin{equation}
\omega(\phi)+{3\over 2}= { {A \phi^2}\over \left(\phi-\phi_0\right)^2},
\label{jesus10}
\end{equation}
where $A$ is an arbitrary positive constant. This form for
$\omega(\phi)$ was first considered by Mimoso and Wands
\cite{Mimoso1995a} (in a theory without a potential). Now, we obtain
\begin{equation}
\phi=\phi_0+B\, e^{\mp {\varphi\over \sqrt{A}}},
\label{jesus11}
\end{equation}
where $B$ is a constant, and the potential, defined by equation
\eref{Vdefine}, is given by
\begin{equation}
V(\phi)=V_0\,\phi^2\left(\phi-\phi_0\right)^{\mp\sqrt{A}k}.
\label{jesus12}
\end{equation}
As before, at the equilibrium points we can express $\phi$ as a
function of $t^{*}$, which then allows us to compute $t$ as a function of
$t^{*}$. At late times we find that
\begin{equation}
\phi\sim \phi_0+t^\beta,
\label{jesus13}
\end{equation}
where $\beta$ is a constant whose sign depends on $k$, $\omega_0$ and
the choice of one of the signs in the theory. What is important
here is that in this case, at late times, we find that the scalar
field tends to a constant value for $\beta<0$, thereby yielding a GR
limit.  In both of the examples considered above, the conformal
transformation for the equilibrium points is regular.

Of course, these are not the only possible forms for a variable
$\omega(\phi)$.  For example, Barrow and Mimoso \cite{Barrow1994a}
studied models with $2\omega(\phi)+3 \propto \phi^\alpha$ $(\alpha>0)$
satisfying the GR limit asymptotically.  (The GR limit
is only obtained asymptotically as $\phi\rightarrow \infty$, although
for a finite but large value of $\phi$ the theory can have a limit
which is as close to GR as is required).  However, by studying the
evolution of the gravitational ``constant'' $G$ from the full Einstein
field equations (i.e., not just the weak-field approximation),
Nordtvedt \cite{Nordtvedt1970a,Nordtvedt1968b} has shown that
\be
\frac{\dot{G}}{G} = -\left( \frac{3+2\omega}{4+2\omega} \right)
\left( 1+\frac{2 \omega'}{(3+2\omega)^2} \right),
\ee
where $\omega' = d\omega/d\phi$
(so that the correct GR limit is only obtained as $\omega\rightarrow
\infty$ and $\omega' \omega^{-3}\rightarrow 0$).  Torres
\cite{Torres1995a} showed that when $2\omega(\phi) + 3 \propto \phi^\alpha$,
$G(t)$ decreases logarithmically and hence $G\rightarrow 0$
asymptotically.  In the above work, no potential was included.  For a
theory with $2\omega(\phi) +3 \propto \phi^\alpha$ and with a non-zero
potential satisfying equation \eref{Vdefine} we have that
\be
\frac{\phi}{V}\frac{dV}{d\phi} = A + B \phi^\alpha
\ee
($\alpha\neq 0$; $A$ and $B$ constants), so that 
\be
V(\phi) = V_0 \phi^A e^{B\phi^\alpha/\alpha}.
\ee
A potential of this form was considered by Barrow \cite{Barrow1993c}.

Finally, Barrow and Parsons \cite{Barrow1997d} have studied three
parameterized classes of models for $\omega(\phi)$ which permit
$\omega\rightarrow\infty$ as $\phi\rightarrow\phi_0$ (where the
constant $\phi_0$ can be taken as $\phi$ evaluated at the present
time) and hence have an appropriate GR limit;
\begin{eqnarray*}
(i) & & 2 \omega(\phi) + 3 = 2 B_1^2 \left| 1-\phi/\phi_0 \right|^{-\alpha} 
\qquad (\alpha>\half), \\
(ii) & & 2\omega(\phi) + 3 = B_2^2 \left| ln(\phi/\phi_0) \right|^{-2| \delta |}
\qquad (\delta>\half), \\
(iii) && 2\omega(\phi) + 3 = B_3^2 \left| 1-(\phi/\phi_0)^{|\beta|} \right|^{-1}
\qquad (\forall~\beta).
\end{eqnarray*}
Other possible forms for $\omega(\phi)$ were discussed in Barrow and
Carr \cite{Barrow1996a} and, in particular, they considered models
$(i)$ above but allowed $\alpha<0$ in order for a possible GR limit to
be obtained also as $\phi\rightarrow \infty$.  Schwinger
\cite{Schwinger1970a} has suggested the form
$2\omega(\phi)+3=B^2/\phi$ based on physical considerations.

\section{\bf Applications}

Let us exploit the formal equivalence of the class of scalar-tensor
theories \eref{bdaction} with $\omega(\phi)$ and $V(\phi)$ given by
\benumber
\omega(\phi) = \omega_0, \qquad V(\phi)=\beta\phi^\alpha, \label{omega}
\eenumber 
with that of GR containing a scalar field
and an exponential potential \eref{exp}.  Indeed, since the conformal
transformation \eref{trans1} is well-defined in all cases of interest,
the Bianchi type is invariant under the transformation and we can
deduce the asymptotic properties of the scalar-tensor theories from
the corresponding behaviour in the Einstein frame.  Also, we have that
\benumber k\equiv\frac{\alpha-2}{\bar{\omega}}, 
\qquad \bar{\omega}^2\equiv \omega_0
+\frac{3}{2}. \label{kval} \eenumber

We recall that at the finite equilibrium points in the Einstein frame
we have that
\begin{eqnarray}
\theta^* & =  & \theta^*_0t_*^{-1}, \label{Etheta} \\
\varphi(t^*) & = & \varphi_0 - \frac{2}{k}\ln(t^*), \label{Ephi}
\end{eqnarray} where \benumber \theta^*_0 = 1+\frac{k^2}{2} e^{k\varphi_0}. 
\eenumber
Integrating equation \eref{trans2} we obtain
\benumber
\phi(t^*) = d \exp{\left(\bar{\omega}^{-1}\varphi(t^*)\right)} = \phi_0 t_* 
^{-2/k\bar{\omega}}, \label{Jphi} \eenumber 
where the constant $\phi_0 \equiv d \exp(\varphi_0/\bar{\omega})$ and
we recall that $t$ and $t^*$ are related by equation \eref{jesus2}, 
and equation \eref{trans1} can be written as
\benumber
g_{ab}=\phi^{-1} g^*_{ab}. \label{intrans1}
\eenumber

\subsection{\bf Examples}
{\bf 1) \ \ }
All initially expanding scalar field Bianchi models with an
exponential potential \eref{exp} with $0<k^2<2$ within general
relativity (except for a subclass of models of type IX) isotropize to the
future towards the power-law inflationary flat FRW model
\cite{Kitada1992a}, whose metric is given by
\benumber
ds^2 = -dt_*^2 + t_*^{4/k^2}\left( dx^2 + dy^2 + dz^2 \right). \label{powerlaw}
\eenumber 
In the scalar-tensor theory (in the Jordan frame), $\phi$ is given by equation 
\eref{Jphi} and from \eref{intrans1} we have that
\benumber
ds_{ST}^2 = \phi_0^{-1} t_*^{2/k\bar{\omega}} \left\{ds^2\right\}.
\label{conform} \eenumber
Defining a new time coordinate by 
\benumber T=ct_*^{\frac{1+k\bar{\omega}}{k\bar{\omega}}}; \qquad c\equiv 
\frac{k\bar{\omega}}{1+k\bar{\omega}} \phi_0^{-\half} \label{newT} \eenumber
(where $k\bar{\omega}+1 \neq 0$; i.e., $\alpha\neq 1$), we obtain
after a constant rescaling of the spatial coordinates
\benumber
ds_{ST}^2 = -dT^2+T^{2K}\left(dX^2+dY^2+dZ^2\right), \label{newds}
\eenumber
where
\be
K\equiv \frac{k^2+2k\bar{\omega}}{k^2(1+k\bar{\omega})}.
\ee
Finally, the scalar field is given by 
\benumber
\phi=\phi_0 c^\frac{2}{1+k\bar{\omega}} T^\frac{-2}{1+k\bar{\omega}} 
= \bar{\phi_0} T^\frac{2}{1-\alpha}. \label{newphi}
\eenumber

Therefore, all initally-expanding spatially-homogeneous models in 
scalar-tensor theories obeying
\eref{omega} with $0<(\alpha-2)^2 < 2\omega_0+3$ (except for a
subclass of Bianchi IX models which recollapse) will asymptote towards
the exact power-law flat FRW model given by equations \eref{newds}
and \eref{newphi}, which will always be inflationary since
$K=\frac{1+\alpha+2\omega_0}{(\alpha-1)(\alpha-2)} > 1$ [note that
whenever $2\omega_0 > (\alpha-2)^2-3 = \alpha^2-4\alpha + 1$, we have
that $1+\alpha+2\omega_0 > \alpha^2-3\alpha+2 =(\alpha-1)(\alpha-2)$].

When $k^2>2$, the models in the Einstein frame cannot inflate and may
or may not isotropize.  Let us consider two examples.

{\bf 2) \ \ }  Scalar field models of Bianchi type VI$_h$ with an exponential potential \eref{exp}  with $k^2 > 2$ asymptote to the future towards the anisotropic Feinstein-Ib\'{a}\~{n}ez model \cite{Feinstein1993a} given by ($m\neq 1$)
\benumber
ds^2=-dt_*^2 + a_0^2 \left(t_*^{2p_1} dx^2 + t_*^{2p_2} e^{2mx} dy^2 + 
t_*^{2p_3} e^{2x} dz^2 \right), 
\eenumber
where the constants obey
\begin{eqnarray}\nonumber
p_1 & = & 1, \\ 
p_2 & = & \frac{2}{k^2} \left(1 + \frac{(k^2-2)(m^2+m)}{2(m^2+1)} \right), \\
\nonumber
p_3 & = & \frac{2}{k^2} \left(1 + \frac{(k^2-2)(m+1)}{2(m^2+1)} \right).
\end{eqnarray}

In the scalar-tensor theory (in the Jordan frame), $\phi$ is given by equation
\eref{Jphi} and the metric is given by \eref{conform}.  After defining the new
time coordinate given by \eref{newT}, we obtain 
\benumber
ds_{ST}^2 = -dT^2 +A_0^2 \left(T^{2q_1} dX^2 + T^{2q_2} e^{2mX}dY^2 + T^{2q_3}
e^{2X} dZ^2 
\right), \label{finds}
\eenumber
where
\benumber q_i \equiv \frac{1+k\bar{\omega}p_i}{1+k\bar{\omega}} ~~~~(i=1,2,3); 
\qquad A_0^2 =a_0^2 \phi_0^{-1}c^{-2q_1}, \eenumber 
and $Y$ and $Z$ are obtained by a simple constant rescaling (and
 $X=x$).  Finally, the scalar field is given by equation
 \eref{newphi}.

The corresponding exact Bianchi VI$_h$ scalar-tensor theory solution
is therefore given by equations \eref{newphi} and \eref{finds} in the
coordinates $(T,X,Y,Z)$.  Consequently, all Bianchi type VI$_h$ models in
the scalar-tensor theory satisfying equations \eref{omega} with
$(\alpha-2)^2 > 2\omega_0 + 3$ asymptote towards the exact anisotropic
solution given by equations \eref{newphi} and \eref{finds}.

{\bf 3) \ \ } An open set of scalar field models of Bianchi type
VII$_h$ with an exponential potential with $k^2>2$ asymptote towards
the isotropic (but non-inflationary) negative-curvature FRW model
\cite{Coley1997a} with metric
\benumber
ds^2 = -dt_*^2 + t_*^2 d\sigma^2, \label{ex3}
\eenumber
where $d\sigma^2$ is the three-metric of a space of constant negative 
curvature.  Again, $\phi$ is given by \eref{Jphi} and the metric is
given by \eref{conform}, which becomes after the time
recoordinatization \eref{newT}
\benumber
ds_{ST}^2 = -dT^2 +C^2T^2 d\sigma^2, \label{ex4}
\eenumber
where $C^2\equiv \phi_0^{-1}c^{-2} = \left[\frac{1+k\bar{\omega}}
{k\bar{\omega}}\right]^2$.  This negatively-curved FRW metric is
equivalent to that given by \eref{ex3}.  Finally, the scalar field is
given by equation \eref{newphi}.

Therefore, when $(\alpha-2)^2> 2\omega_0+3$, there is an open set of
(BVII$_h$) scalar-tensor theory solutions satisfying equations
\eref{omega} which asymptote towards the exact isotropic solution
given by equations \eref{newphi} and \eref{ex4}.

Equations \eref{Ephi} and \eref{Jphi} and the resulting analysis are
only valid for scalar-tensor theories satisfying \eref{omega}.
However, the asymptotic analysis will also apply to generalized
theories of the forms discussed in subsections 2.2.1 and 2.2.2.
Finally, a similar analysis can be applied in Brans-Dicke theory with $V=0$
\cite{Coley1998a}.

\subsection{\bf Self-Similarity}

All three attracting scalar-tensor theory solutions in
the last subsection are self-similar; metric \eref{newds} admits the
homothetic vector (HV) $\bv{X} = T\bv{\frac{\di}{\di T}} + (1-K)\left\{
X\bv{\frac{\di}{\di X}} +Y\bv{\frac{\di}{\di Y}}+Z\bv{\frac{\di}{\di Z}} \right\}$,
metric \eref{finds} admits the HV $\bv{X} = T\bv{\frac{\di}{\di T}} + (1-q_2)
Y\bv{\frac{\di}{\di Y}}+(1-q_3)Z\bv{\frac{\di}{\di Z}} $, and metric \eref{ex4} admits the HV
$\bv{X}=T\bv{\frac{\di}{\di T}}$.  Of course, all three solutions in the
corresponding general relativistic model (i.e., in the Einstein frame)
are self-similar.  Let us show that this is always the case;
i.e., all scalar-tensor solutions obtained in this way are
self-similar.

In \cite{Coley1997a} it was shown that the cosmological solutions
corresponding to the finite equilibrium points of the ``reduced
dynamical system'' of the spatially homogeneous scalar field models
with an exponential potential are all self-similar.  Let $g^*_{ab}$ be
the metric of such a solution and $\bv{X}_*$ the corresponding HV;
hence we have that
\benumber
{\cal L}_{\bv{X}_*}g^*_{ab}=2g^*_{ab}, \label{Lie}
\eenumber
where ${\cal L}$ denotes the Lie derivative along $\bv{X}_*$.  In the
coordinates in which $\theta^*=\theta^*_0t_*^{-1}$, from ${\cal
L}_{\bf{X}_*} \theta^* =-\theta^*$ we find that \cite{Coley1998a}
\benumber
\bv{X}_* = t^*\frac{\di}{\di t^*} + X^\mu_*(x^\gamma_*)\frac{\di}{\di 
x^\mu_*}.
\eenumber

Now, the metric $g_{ab}$ in the corresponding scalar-tensor theory is given
by \eref{intrans1}, where the scalar field is given by \eref{Jphi}, viz., 
\benumber
\phi(t^*) = \phi_0 t^{-2/k\bar{\omega}}_* \label{phifin}
\eenumber
(or by \eref{newphi} in terms of the time coordinate $T$).  We emphasize
that this {\em {power-law}} form for $\phi$ is only valid for
scalar-tensor theories that obey conditions \eref{omega}.  Hence, from 
equations \eref{Lie}-\eref{phifin} it follows that
\begin{eqnarray}
\nonumber
{\cal L}_{\bv{X}_*}g_{ab} & = & \bv{X}_*\left(\phi^{-1}(t^*)\right) g^*_{ab} 
   + \phi^{-1} {\cal L}_{\bv{X}_*} g^*_{ab} \\
\nonumber   & = & t^* \frac{\di}{\di t^*} \left(\phi^{-1}_0  
t_*^\frac{2}{k\bar{\omega}} \right) g^*_{ab} + 2\phi_0^{-1} t_*^\frac{2}{k\bar{\omega}} g^*_{ab} \\
\nonumber & = & \left(\frac{2}{k\bar{\omega}}+2\right) \phi^{-1}_0  
t_*^\frac{2}{k\bar{\omega}}  g^*_{ab} \\
 & = & 2c g_{ab}, \label{Liedone}
\end{eqnarray}
where the constant c is given by
$c=(1+k\bar{\omega})/(k\bar{\omega})$.  That is, $\bv{X}=\bv{X}_*$ is
a homothetic vector for the spacetime with metric $g_{ab}$ and
consequently the corresponding scalar-tensor theory solution is
self-similar.

\subsection{\bf The Special Case $\alpha=1$}

In the analysis above we have omitted the special case $\alpha=1$
(i.e., $k\bar{\omega}=-1$).  This case is degenerate
as we will now demonstrate.  Let the general relativistic metric be
defined by
\benumber
ds^2=-dt_*^2+\gamma_{\mu \nu} dx^\mu dx^\nu. \label{Egen}
\eenumber
First, suppose we take $k\bar{\omega}=-1$ in \eref{Jphi} and define a
new time coordinate by
\benumber
T=\phi_0^{-\half} \ln(t_*),
\eenumber
then the metric \eref{Egen} becomes
\benumber
ds_{ST}^2=-dT^2+\phi_0^{-1}\exp{\left(-2\sqrt{\phi_0}
T\right)}\gamma_{\mu \nu} dx^\mu dx^\nu,
\label{dsa}
\eenumber
where $\phi(T)= \phi_0 \exp{\left(2\sqrt{\phi_0}T\right)}$.  Now, from
equation
\eref{Liedone} we obtain
\benumber
{\cal L}_{\bv{X}_*} g_{ab}=0;
\eenumber
i.e., in this case $\bv{X}=\bv{X}_*$ is a Killing vector (KV) of the
spacetime \eref{dsa}.
Since the KV $\bv{X}$ is timelike, the spatially homogeneous metric
\eref{dsa} admits four KV acting simply transitively and hence the
resulting spacetime is (totally - i.e., four-dimensionally) homogeneous.

All known non-flat homogeneous spacetimes are given in table 10.1 in
\cite{Kramer1980a}; hence the metric \eref{dsa} is given by one of
those spacetimes in this table representing an 
orthogonal spatially homogeneous metric with a diagonal Einstein
tensor (representing a perfect fluid spacetime or an Einstein
spacetime with a cosmological constant) - all of these metrics are
indeed known \cite{Kramer1980a}.  In the case when metric \eref{dsa}
is the flat Minkowski metric, the corresponding general relativistic
spacetime \eref{Egen} is de Sitter spacetime.  However, this
corresponds to the degenerate case in which
\be
\theta^*=\theta^*_0, ~~~\mbox{a constant};
\ee
this is the only possibility in which equation \eref{Etheta} is not
valid and hence the resulting analysis does not follow.  This
degenerate case corresponds to $k=0$ in \eref{exp} (i.e., $V^* = V_0$,
a constant); since $k\bar{\omega}=-1$ this corresponds to
$\bar{\omega}\rightarrow \infty$ or $\omega_0\rightarrow \infty$ (in
which case GR is recovered from the scalar-tensor theory under
consideration).

Finally, if $\alpha=1$ in \eref{power} (i.e., $V=\beta\phi$), then the
action \eref{bdaction} becomes
\be
S=\int \sqrt{-g} \left[ \phi (R-2\beta) - \frac{\omega_0}{\phi}g^{ab}
\phi_{,a}\phi_{,b} + 2{\cal{L}}_m \right] d^4x, \ee
which is equivalent to that for Brans-Dicke theory incorporating an
additional constant $\beta$.  Under the conformal transformation and
field redefinition \eref{trans} the action becomes that for general
relativity with a cosmological constant (and additional matter
fields), and from the cosmic no-hair theorem it follows that all
spatially homogeneous models (except for a subclass of Bianchi type IX)
asymptote to the future towards the de Sitter model \cite{Wald1983a}.

\section{\bf Conclusions}

In this paper we have studied the asymptotic behaviour of a special
subclass of spatially homogeneous cosmological models in scalar-tensor
theories which are conformally equivalent to general relativistic
Bianchi models containing a scalar field with an exponential potential
by exploiting results found in previous work \cite{Coley1997a}.

We illustrated the method by studying the particular example of
Brans-Dicke theory with a power-law potential and various
generalizations thereof, paying particular attention to the possible
isotropization and inflation of such models.  In addition, we
discussed physical constraints on possible late-time behaviour and, in
particular, whether the scalar-tensor theories under consideration
have a general relativistic limit at late times.

In particular, several exact scalar-tensor theory cosmological models
(both inflationary and non-inflationary, isotropic and anisotropic)
which act as attractors were discussed, and all such exact
scalar-tensor solutions were shown to be self-similar.  

This is related to the previous work of several authors.
Specifically, Chauvet and Cer\-vantes-Cota \cite{Chauvet1995a} studied
isotropization in Brans-Dicke gravity including a perfect fluid
with $p=(\gamma-1)\mu$.  They examined whether the anisotropic models
contain an FRW model as an asymptotic limit, which is how they defined
isotropization.  For Bianchi models of types $I$, $V$ and $IX$, they
found exact solutions in these cosmologies which can isotropize to the
future, depending on the values of $\gamma$ and $\omega$ and two other
arbitrary constants.  Furthermore, Mimoso and Wands \cite{Mimoso1995a}
also studied scalar-tensor models with variable $\omega$ without a
self-interacting potential $V$ but coupled to barotropic matter.
Regarding the possible isotropization of the cosmological models (meaning 
here that the shear of the fluid becomes negligible), they
concentrated on models of Bianchi type $I$ and first discussed
constraints on a fixed $\omega=\omega_0$ model necessary for isotropization at
late times.  In the particular case of a false vacuum ($p=-\mu$), they
showed that the de Sitter solution is the late-time  attractor of the model.
They then proceeded to examine arbitrary $\omega(\phi)$ Bianchi type
$I$ cosmologies and showed that if a solution is to asymptote
towards a GR limit (i.e., $\omega\rightarrow\infty$) then it must also
isotropize.  Their paper also discussed initial singularities in
models of other Bianchi types.

The work in this paper can be generalized in a number of ways.  In
particular, more general scalar-tensor theories can be considered and
more general (than spatially homogeneous) geometries can be studied.
For example, the more general class of inhomogeneous $G_2$ models could be
considered \cite{Ibanez1996a,Feinstein1995a,Aguirregabiria1993a} in
which there exists two commuting spacelike Killing vectors.  The
motivation for studying $G_2$ cosmologies is that there is some
evidence that the class of {\em self-similar} $G_2$ models plays an
important r\^{o}le in describing the asymptotic behaviour of more
generic general relativistic scalar field models with an exponential potential
[cf. 42]; in this way, we may be able to find special 
scalar-tensor $G_2$ cosmological models that describe the asymptotic
properties of more general scalar-tensor cosmologies.  Some potential
problems that exist in this more general context is that since $\phi$,
and hence the transformation \eref{trans1}, depends on both time and
one space variable, the transformation \eref{trans} will be singular
(at least for certain values of the space variable) and the
classification of $G_2$ models may not be preserved under such a
transformation.

\noindent{\bf {Acknowledgments}}

We would like to thank Robert van den Hoogen and Itsaso Olasagasti for
helpful comments.  This work was funded, in part, by NSERC (APB and
AAC), the I.W. Killam Fund (APB) and CICYT PB96-0250 (JI).

\end{document}